\documentclass[10pt]{article}
\usepackage{psfig,epsf,conf-X}

\begin{document} 
\small
\heading{%
%Begin Headingthere is   

%
Galaxy Cluster Shapes
% End Heading
}
\par\medskip\noindent
\author{%
%Begin Author names
Spyros Basilakos $^{1,2}$, Manolis Plionis $^1$ \& Steve Maddox $^3$
%End Author names
}
\address{%
%First address
Institute of Astronomy \& Astrophysics, National Observatory of Athens, 
I.Metaxa \& B.Pavlou, Palaia Penteli, 152 36 Athens, Greece
}
\address{%
% Second Address
Astrophysics Group, Imperial College London, Blackett Laboratory, 
Prince Consort Road, London SW7 2BZ, UK
}
\address{%
% Third Address
School of Physics and Astronomy, University of Nottingham,
Nottingham NG7 2RD, UK
}

\begin{abstract}
We estimate the distribution 
of intrinsic shapes of the APM galaxy clusters from their corresponding 
distribution of apparent shapes.
We smooth the discrete galaxy distribution and
define the cluster shape by fitting the best ellipse to the different 
isodensity contours.
Using Monte-Carlo simulations we have studied the performance of our method 
in the presence of the expected galaxy background, at the different distances 
traced by the APM clusters, and we have devised a method to correct
statistically the projected cluster shapes for discreteness effects and random
fluctuations. We find that the true cluster shape is consistent with that
of prolate spheroids.
\end{abstract}

\section{Method}
In order to estimate the projected cluster shape
we diagonalize the inertia tensor (${\rm det} (I_{ij} - 
\lambda^2 M_{2})=0$) where $M_{2}$ is the $2 \times 2$ unit 
matrix. The eigenvalues $(\lambda_1, \lambda_2)$ with $(\lambda_2>\lambda_1)$ 
define the ellipticity of the configuration under study: 
$\varepsilon=1-\lambda_1/\lambda_2$.
Initially the galaxy positions are transformed to the coordinate system of 
each cluster. Then the discrete galaxy distribution is smoothed 
using a Gaussian kernel. All cells that have a density above some 
threshold are used to define the moments of inertia with weight
$w_{i}= (\rho_{i}-\langle \rho \rangle) / \langle \rho \rangle$
where $\langle \rho \rangle$ is the mean projected APM galaxy density.
This method is free of the aperture bias and we found that it performs
significantly better than using the discrete galaxy distribution.

\section{Results}
%The crosses in figure 1 show the projected axial ratio distributions
%with the Poisson 1$\sigma$ error bars and the solid lines show the
%kernel estimate $\hat{f}$ (cf. \cite{Ryden}).
%\begin{figure}
%\mbox{\epsfxsize=8cm \epsffile{freq.ps}}
%\end{figure}
%In the top panel we present our results for the uncorrected
%distribution of all APM clusters; while 
%in the bottom panel we present the corrected distribution using
%the 405 APM clusters free of significant subclustering.
%It is obvious that the two distributions are significantly different,
%with the peak of the corrected distribution having moved to lower $q$'s but
%with an extended contribution of apparently quasi-spherical objects.
Inverting a set of integral equations, which relate the projected and real
axial ratio distribution, we obtain the distribution of real axial ratios
under the assumption that the orientations are random with 
respect of line of sight.
\begin{figure}
\mbox{\epsfxsize=8.5cm \epsffile{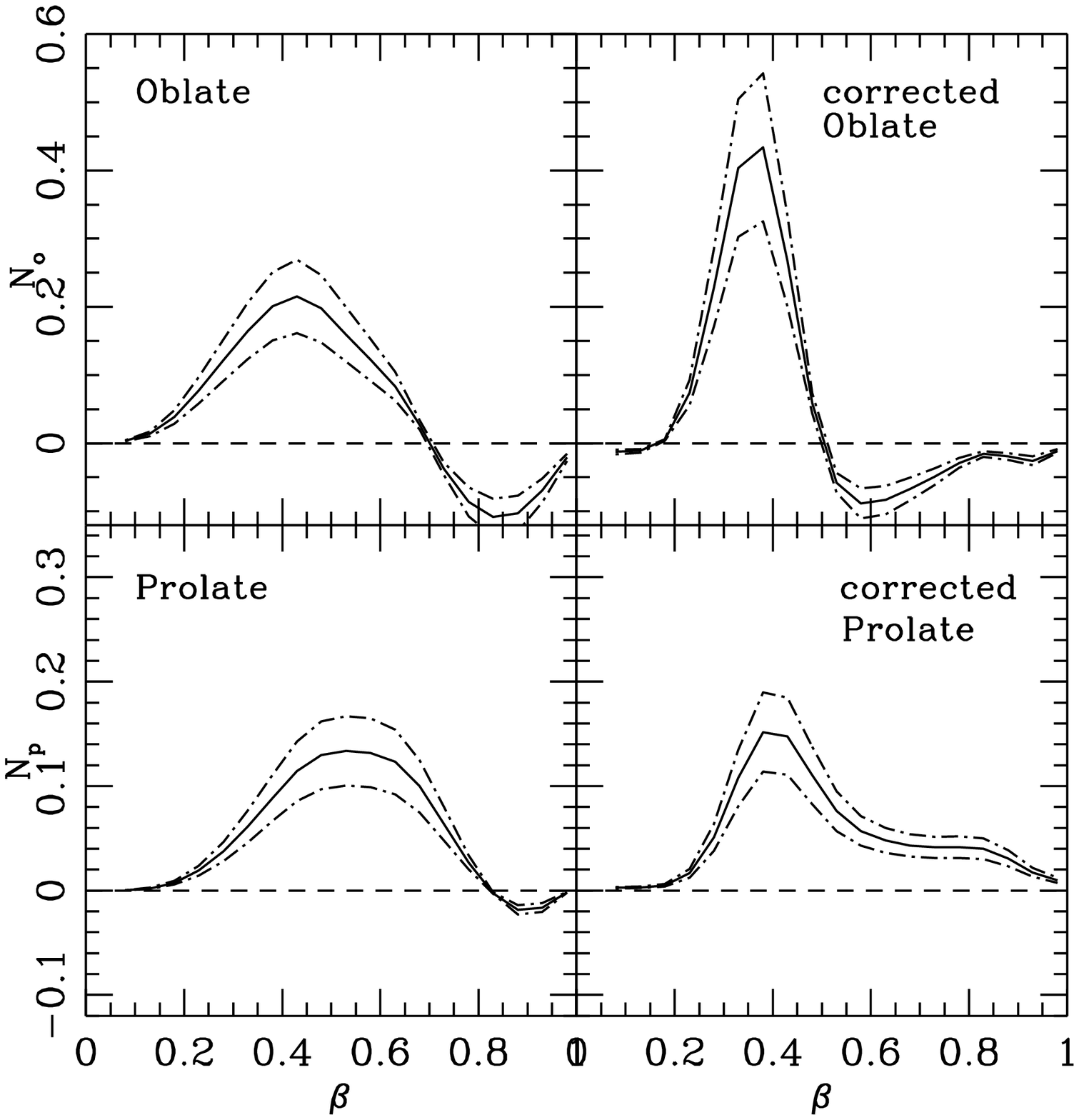}}
\end{figure}
According to \cite{Ryden}, if the inverted distribution of axial ratios 
has significantly negative values, a fact which is unphysical, 
then this can be viewed 
as a strong indication that the particular spheroidal model is unacceptable.
In figure 1 we present the uncorrected and corrected intrinsic axial
ratio distributions. It is evident that the APM cluster shapes are 
better represented by that the prolate spheroids (in agreement
with \cite{Plionis}) rather than oblate beacause
the former model provides a distribution
of intrinsic axial ratios that is positive over the whole axial ratio
range.

\end{document}